\documentclass[aps, prl, twocolumn, showkeys]{revtex4-1}

\usepackage{chemformula} 
\usepackage{siunitx} 
\usepackage[inline,shortlabels]{enumitem} 
\usepackage{hyperref} 
\usepackage{graphicx} 
\usepackage{tabularx}
\usepackage[referable]{threeparttablex}

\usepackage{natbib}

\makeatletter
\newcommand{\thickhline}[1]{%
    \noalign {\ifnum 0=`}\fi \hrule height #1
    \futurelet \reserved@a \@xhline
}
\makeatother

\begin{document}

    \title{Magnetic, electrochemical and thermoelectric properties of P2 - \ch{Na_x(Co_{7/8}Sb_{1/8})O2}}
    \author{M. H. N. Assadi}
    \affiliation{Center for Computational Sciences, University of Tsukuba, Tsukuba, Ibaraki 305-8577, Japan.}
    \email{h.assadi.2008@ieee.org}
    \altaffiliation[Tel:]{+81-29-860-4709}
    \author{S. Li}
    \affiliation{School of Materials Science and Engineering, University of New South Wales, Sydney, NSW 2052, Australia.}
    \author{R. K. Zheng}
    \affiliation{School of Physics, University of Sydney, Sydney, New South Wales 2006, Australia.}
    \author{S. P. Ringer}
    \affiliation{Australian Institute for Nanoscale Science and Technology, The University of Sydney, NSW, 2006, Australia.}
    \author{A. B. Yu}
    \affiliation{Department of Chemical Engineering, Monash University, Clayton, Vic. 3800, Australia.}
    \date{2017}

    \begin{abstract}
        We theoretically investigated the electronic, electrochemical and magnetic properties of Sb doped \ch{Na_{x}CoO2}  ($x = 1, 0.75$ and 0.50). \ch{Sb_{Co}} dopants adopt +5 oxidation state in \ch{Na_{x}CoO2}  host lattice for all Na concentrations (x). Due to high oxidation states, \ch{Sb^{5+}} strongly repels Na ions and therefore it decreases the electrochemical potential (vs. Na/\ch{Na^{+}}). The electrons introduced by \ch{Sb^{5+}} localize on nearby Co ions creating \ch{Co^{2+}} species which are absent in undoped \ch{Na_{x}CoO2}. \ch{Co^{2+}} ions reduce the spin entropy flow decreasing the Seebeck coefficient in the Sb doped compounds. The results can be generalized to other dopants with high oxidation state.
    \end{abstract}
    \maketitle

    \section{Introduction}
\begin{figure}
            \centering
            \includegraphics[width=0.9\columnwidth]{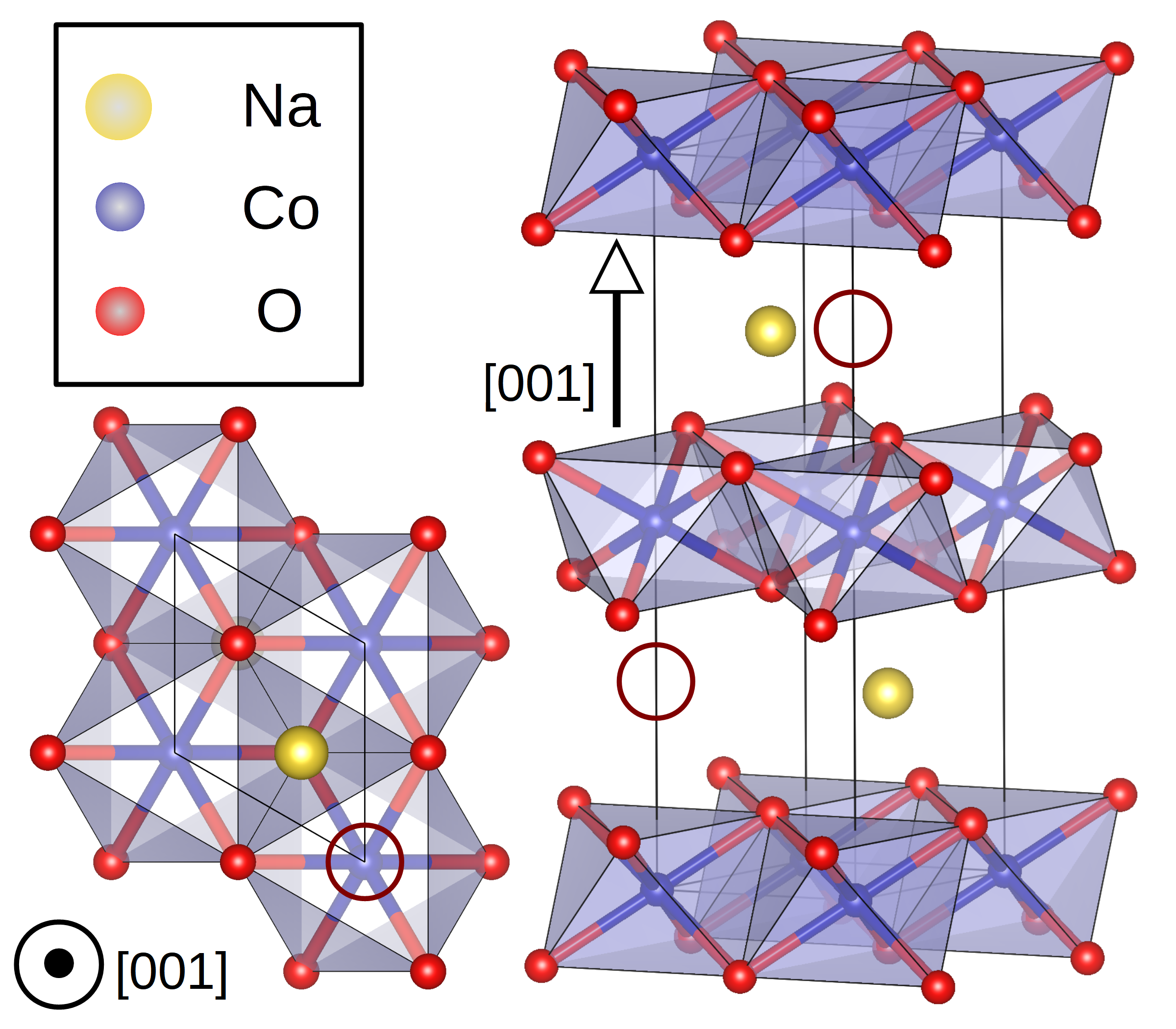}
            \caption{\label{fig:1}The top and side views of \ch{Na_{1}CoO2}  P63/mmc crystal structure. Co ions occupy 2a sites (0,0,0), O ions occupy 4f sites (1/3, 2/3, z). In \ch{Na_{1}CoO2}, all Na ions occupy Na2 sites which are equivalent to 2d Wyckoff position (2/3, 1/3, $\tfrac{1}{4}$). For smaller Na content, some Na ions occupy Na1 sites with 2b Wyckoff representation (0, 0, $\tfrac{1}{4}$). 2b site are marked by brown circles for clarity.}
        \end{figure}

        Sodium cobaltate (\ch{Na_{x}CoO2}) is an interesting and promising compound for high efficiency thermoelectric \cite{Fergus2012} and sodium ion battery applications \cite{Xiang2015}. This compound also exhibits a rich magnetic and structural phase diagrams \cite{Mendels2005, Berthelot2011}. As demonstrated in FIG. \ref{fig:1}, the primitive cell of P2-\ch{Na_{x}CoO2}  lattice is made of two alternating Na layers and edge-sharing \ch{CoO6} octahedra. Due to its frustrated triangular nature, the \ch{CoO2} layer in \ch{Na_{x}CoO2}  creates a large spin entropy flow \cite{Wang2003} and which results in a large Seebeck coefficient \cite{Terasaki1997}. From materials engineering viewpoint, controlling Na concentration ($x$) has been the primary technique to tailor the properties of \ch{Na_{x}CoO2}  for the desired applications. However, given the volatile nature of Na ions \cite{Cushing1999}, establishing the precise property-structure relationship experimentally is somehow challenging. This experimental difficulty has motivated ongoing theoretical investigations into the structural and electronic properties of pristine \ch{Na_{x}CoO2}  \cite{Zhang2005, Meng2008, Wang2007, Wang2008}.

        Doping \ch{Na_{x}CoO2}  with other elements is another plausible method to improve the sodium cobaltate's thermoelectric \cite{Nagira2004, Seetawan2006} and electrochemical properties \cite{Dolle2005, Bo2016}. Nonetheless, further progress in realizing high performance functional applications of the doped \ch{Na_{x}CoO2}  compounds requires an accurate understanding of the electronic structure of the doped P2-\ch{Na_{x}CoO2}. However, due to complex crystal structure of and strong electronic correlation in Na deficient sodium cobaltate, the theoretical aspects of doped \ch{Na_{x}CoO2}  have not been fully explored. Accordingly, in this work, we study how Co's substitution with Sb influences the electronic behavior of \ch{Na_{x}CoO2}. Sb doping was particularly chosen as it has recently shown to improve the electrochemical performance of other layered cathode materials \cite{Twu2015, Twu2017}. Some of the conclusions drawn here can be applied to predict the magnetic, electrochemical and thermoelectric properties of \ch{Na_xCO2} compounds in which dopants with high oxidation state partially replace Co.

    \section{Computational and System Settings}
        Spin-polarized density functional calculations based on the projector augmented wave method \cite{Blochl1994, Kresse1999} were performed with VASP \cite{Kresse1996a, Kresse1996b}. Generalized gradient approximation (GGA) based on Perdew-Wang formalism \cite{Perdew1996, Perdew1997} was applied to approximate the exchange-correlation functional. The energy cutoff was set to be $500  \si{.eV}$ . Brillouin zone sampling for the supercells was carried out by choosing a $k$-point grid with ~$0.05\si{.\angstrom^{-1}}$ spacing between the $k$ points generated by Monkhorst-Pack scheme \cite{Monkhorst1976}. Orbital projected density of states (DOS) were calculated using LOBSTER package \cite{Maintz2013, Maintz2016}. To improve accuracy, we added on-site Coulomb ($U$) and exchange ($J$) interaction terms of $U = 6$ and $J  = 1\si{.eV}$ ($U\ch{_{eff}}  = 5\si{.eV}$) to Co 3d elections using Dudarev's approach \cite{Dudarev1998}. Among many values reported for $U$ and $J$ in the literature \cite{Lee2006, Hinuma2008, Molenda2014}, the chosen $U$\ch{_{eff}} value reproduces the charge disproportionation of the Co ions \cite{Mukhamedshin2014, Mukhamedshin2015} and as shown in Table \ref{tab:1}, predicts more accurate electrochemical potential (vs. Na/\ch{Na^{+}}) for \ch{Na_{x}CoO2}. We could nonetheless reproduce the charge disproportionation with smaller $U$\ch{_{eff}} of $4  \si{.eV}$  but this value greatly underestimates the electrochemical potential. Furthermore, for \ch{Na_{0.5}CoO2:Sb}, we compared the accuracy of $U\ch{_{eff}}  = 5\si{.eV}$ against HSE03 functional. Both methods resulted in similar orbital alignment for Co and Sb ions and similar charge densities indicating the adequacy of the chosen $U$\ch{_{eff}}.

        The unit cells of the desodiated \ch{Na_{x}CoO2}  ($x = 0.75$ and 0.5) were constructed by removing Na ions from the pristine \ch{Na_{1}CoO2}  (shown in FIG. \ref{fig:1}). In desodiated \ch{Na_{x}CoO2}, a Na ion can occupy either Na1 site or a Na2 site. The Na1 site is located above the Co ions in \ch{CoO2} layer and the Na2 site is located in above the center of the triangle formed by three Co ions. Na1 and Na2 correspond to 2b and 2d Wyckoff positions respectively in a hexagonal P63/mmc lattice as shown in FIG. \ref{fig:1}. The arrangement of Na in Na1 and Na2 sites depends on Na concertation. In this work, we adopt the Na arrangement obtained in our previous work \cite{Assadi2015b}. Table \ref{tab:1} contains the calculated lattice parameters and the dimensions of the of \ch{Na_{x}CoO2}  unitcell with respect to the \ch{Na_{1}CoO2}  primitive cell. We used a $4a\times2a\times1c$ \ch{Na_{1}CoO2}  and a $2a\times1a\times1c$ \ch{Na_{0.5}CoO2}  supercells and the unitcell of \ch{Na_{0.75}CoO2}  to simulate doped compounds. All doped supercells contained seven Co, one Sb and 16 O ions as described in FIG. \ref{fig:2}. For the sake of clarity, we always substituted a Co from the bottom \ch{CoO2} layer with the Sb dopant. We then relaxed the internal coordinates of all ions in the supercell while fixing the lattice constants to the calculated values of undoped \ch{Na_{x}CoO2}.
       
        \begin{table}
         \centering
        \begin{threeparttable}
        \caption{\label{tab:1}The calculated structural and electrochemical properties of \ch{Na_{x}CoO2}  and their comparison to earlier works. The calculated lattice parameters deviate from the experimental values by less than $1\%$ and the potential (vs. Na/\ch{Na^{+}}) calculated with $U\ch{_{eff}}  = 5\si{.eV}$ deviates from the measurement by less than $3\%$.}
                \begin{tabularx}{\linewidth}{p{2.5cm} c c c} 
            \thickhline{1.5pt}
            Compound & \ch{Na_{1}CoO2} & \ch{Na_{0.75}CoO2}  & \ch{Na_{0.5}CoO2} \\
            \thickhline{1pt}
            $a\left(\mbox{\AA}\right)$(This work) & 2.906 & 2.890 & 2.834\\
            $c\left(\mbox{\AA}\right)$(This work) & 10.566 & 10.899 & 11.171\\
            $a\left(\mbox{\AA}\right)$(Experiment) & 2.8829\tnotex{tn:1a} & 2.8329\tnotex{tn:1b} & 2.81508\tnotex{tn:1a}\\
            $c\left(\mbox{\AA}\right)$(Experiment) & 10.4927\tnotex{tn:1a} & 10.990\tnotex{tn:1b} & 11.1296\tnotex{tn:1a}\\
            Potential vs. Na/\ch{Na+}, ($U\ch{_{eff}} = 5\si{.eV}$) & --- & $2.396 \si{V}$ & $2.880 \si{V}$\\
            Potential vs. Na/\ch{Na+}, ($U\ch{_{eff}} = 4\si{.eV}$) & --- & $1.295 \si{V}$ & $2.501 \si{V}$\\
            Experimental average potential & --- & $2.335 \si{V}$\tnotex{tn:1c} & $2.960 \si{V}$\tnotex{tn:1c}\\
            Unitcell dimension & $1a\times1a\times1c$ & $1a\times\sqrt{3}a\times1c$ & $1a\times2\sqrt{3}a\times1c$\\
            Na1/Na2 (This work) & 0 & 0.5 & 1\\
            Na1/Na2 (Others) & 0\tnotex{tn:1d}\tnote{,}\tnotex{tn:1e} & 0.5\tnotex{tn:1d}\tnote{,}\tnotex{tn:1e} & 1\tnotex{tn:1d}\tnote{,}\tnotex{tn:1e}\\\thickhline{1.5pt}
          \end{tabularx}
          \begin{tablenotes}
        \item [a] \label{tn:1a} \cite{Huang2004}
        \item [b] \label{tn:1b} \cite{Seetawan2006}
        \item [c] \label{tn:1c} \cite{Berthelot2011}
        \item [d] \label{tn:1d} \cite{Zhang2005}
        \item [e] \label{tn:1e} \cite{Meng2008}
        \end{tablenotes}
        \end{threeparttable}
        \end{table}
       
    \section{Results and Discussion}
        It has been established that Sb ions are always stabilized by substituting a Co ions for a wide range of x in the \ch{Na_{x}CoO2}  host lattice \cite{Assadi2015a}. Now, we examine how Sb behave electronically in \ch{Na_{x}CoO2}  host lattice. Sb ions generally adopt two stable oxidation states; \ch{Sb^{5+}} and \ch{Sb^{3+}}. \ch{Sb^{3+}} ion is isovalent to \ch{Co^{3+}} therefore \ch{Co^{3+}} substitution with \ch{Sb^{3+}} does not alter the carrier density in the host system. This means that in \ch{Na_{1}CoO2:Sb^{3+}}, all Co ions are in +3 oxidation state ($t_{2g}^6 e_g^0$). In \ch{Na_{0.75}CoO2:Sb^{3+}}, two \ch{Co^{3+}} convert to \ch{Co^{4+}} ($t_{2g}^5 e_g^0$) to bear the holes generated by Na extraction. Similarly, In \ch{Na_{0.50}CoO2:Sb^{3+}}, four Co ions are in +4 oxidation state. \ch{Sb^{5+}}, on the other hand, introduces two electrons that would be borne on Co ions in an opposite manner to any holes created by Na extraction. Consequently, in \ch{Na_{1}CoO2:Sb^{5+}}, two \ch{Co^{3+}} ($t_{2g}^6 e_g^0$) ions convert to \ch{Co^{2+}} ($t_{2g}^6 e_g^1$). In \ch{Na_{0.75}CoO2:Sb^{5+}}, however, there are two possibilities. First, the two extra electrons generated by \ch{Sb^{5+}} substitution compensate for the two holes created by Na removal leaving all Co ions in \ch{Na_{0.75}CoO2:Sb^{5+}}  at +3 oxidation state ($t_{2g}^6 e_g^0$). The other possibility is that the hole created by Na extraction from the top Na layer localizes on the neighboring Co from the top \ch{CoO2} layer converting a \ch{Co^{3+}} to \ch{Co^{4+}} while simultaneously, the hole created by Na extraction from the bottom \ch{CoO2} layer converts a \ch{Co^{2+}} to back a \ch{Co^{3+}}. In this scenario, there is one \ch{Co^{4+}} ($t_{2g}^5 e_g^0$) in the top \ch{CoO2} layer and one \ch{Co^{2+}} ($t_{2g}^6 e_g^1$) in the bottom \ch{CoO2} layer. In \ch{Na_{0.5}CoO2} \ch{Sb^{5+}}, the two electrons from \ch{Sb^{5+}} convert two nearby \ch{Co^{4+}} ions from the same \ch{CoO2} layer that contains the Sb dopant back to \ch{Co^{3+}}.

        Table \ref{tab:2} shows all different possible oxidation states of Co ions and the relative total energy of \ch{Sb^{3+}} and \ch{Sb^{5+}} doped \ch{Na_{x}CoO2}. For all x values, our results indicate that the \ch{Sb^{5+}} is more stable than \ch{Sb^{3+}} when substituting a Co in the \ch{Na_{x}CoO2}  host lattice. One possible explanation for this trend is that the ionic radius of \ch{Sb^{5+}} in octahedral coordination is $0.60\si{.\angstrom}$ which is closer to the ionic radius of \ch{Co^{3+}} in \ch{Na_{x}CoO2}  which is $0.55\si{.\angstrom}$ \cite{Shannon1976}. The ionic radius of \ch{Sb^{3+}} in octahedral coordination is $0.76\si{.\angstrom}$ which is considerably larger than \ch{Co^{3+}}'s radius therefore its substations would lead to large internal lattice stress. Furthermore, for $x =1.00$ and 0.75, the substitution of Co with Sb adds elections to antibonding Co  $e_g$  states which are $\sim 2.3 \si{.eV}$ higher than the $t_{2g}$ states. In this case, \ch{Sb_{Co}} dopant is, nonetheless, still more stable than Na substituting configurations as the formation energy of the latter is generally 2.2 to $2.8 \si{.eV}$  higher than Co substituting configurations \cite{Assadi2015a}.

\begin{figure*}[!tp]
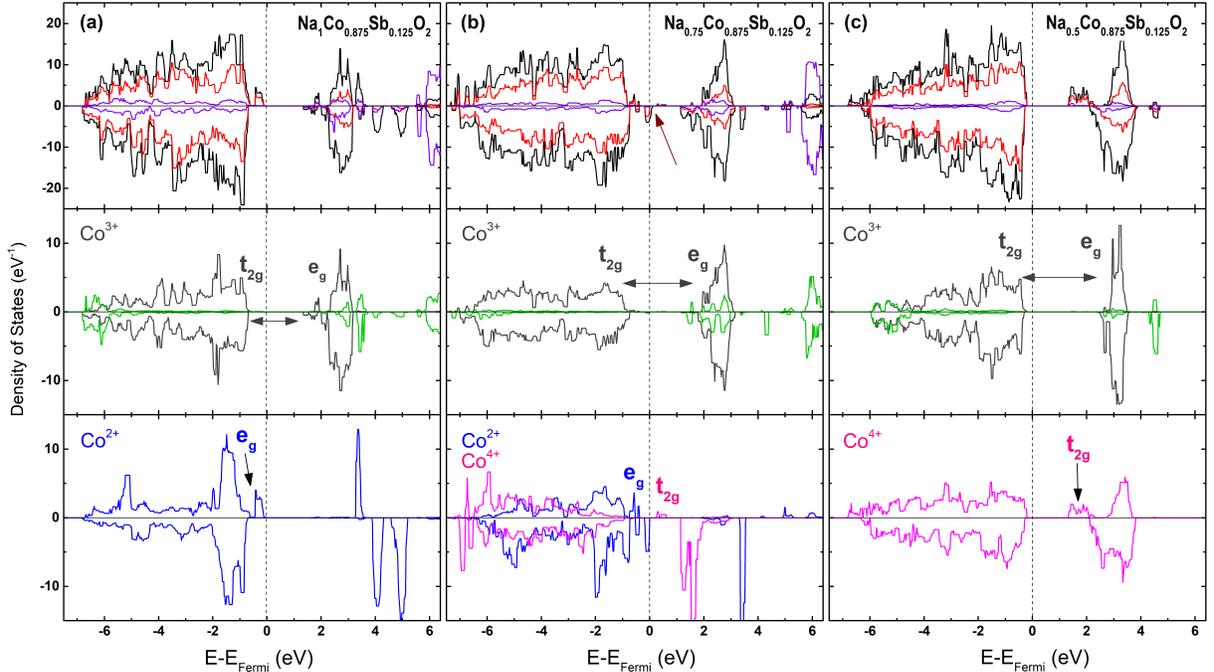

        \begin{minipage}{\textwidth}
            \centering
            \includegraphics[width=0.9\textwidth]{{{Fig3}}}
            \caption{\label{fig:3}Total and partial density of states of the Sb doped \ch{Na_{x}CoO2}  compounds. In the top panels, black, red and purple lines correspond to the total, O 2p, and Na 3s states. In the middle panels, gray and green lines correspond to \ch{Co^{3+}} 3d and Sb states. In the bottom panels, the blue and magenta lines correspond to the \ch{Co^{2+}} 3d and \ch{Co^{4+}} states. Sb, Na, \ch{Co^{2+}} and \ch{Co^{4+}} states were magnified for clarity.}
        \end{minipage}
        \end{figure*}

        FIG. \ref{fig:3} shows the total and partial density of states of \ch{Na_{x}CoO2:Sb^{5+}}.  A common feature among all three Sb doped compounds is the large crystal field splitting among \ch{Co^{3+}} $t_{2g}$ and  $e_g$  states as marked with grey arrows. This prediction agrees with the well-known Co's octahedral splitting of $\sim 2.3 \si{.eV}$ \cite{Roth1964}. Furthermore, the Fermi level of both \ch{Na_{1}CoO2:Sb^{5+}}  and \ch{Na_{0.5}CoO2:Sb^{5+}}  is located in zero DOS region indicating that the Sb doping in these two compounds does not introduce metallic conduction. However, the Fermi level in \ch{Na_{0.75}CoO2}  crosses an  $e_g$  sub-band (marked with a brown arrow in Fig. \ref{fig:3} (b)) that is located near the main valence band creating a metallic conduction in a similar manner to the undoped \ch{Na_{0.75}CoO2}  \cite{Motohashi2003}. Additionally, the filled 5p states gravitate towards the bottom of the valence states indicating Sb's high oxidation state of +5. In \ch{Na_{1}CoO2:Sb^{5+}}  and \ch{Na_{0.75}CoO2:Sb^{5+}}  [FIG. \ref{fig:3}(a) and (b)], \ch{Co^{2+}} $e_g$  states experience further crystal field and ferromagnetic exchange splittings. As a result, filled  $e_g$  states are positioned just above the fully filled \ch{Co^{3+}} $t_{2g}$ states constituting antibonding states while the rest of the  $e_g$  states are located above the Fermi level. In \ch{Na_{0.75}CoO2:Sb^{5+}}  and \ch{Na_{0.5}CoO2:Sb^{5+}}  [FIG. \ref{fig:3}(b) and (c)], on the other hand, all $t_{2g}$ states split and locate above the Fermi level.

        \begin{figure*}[!tp]
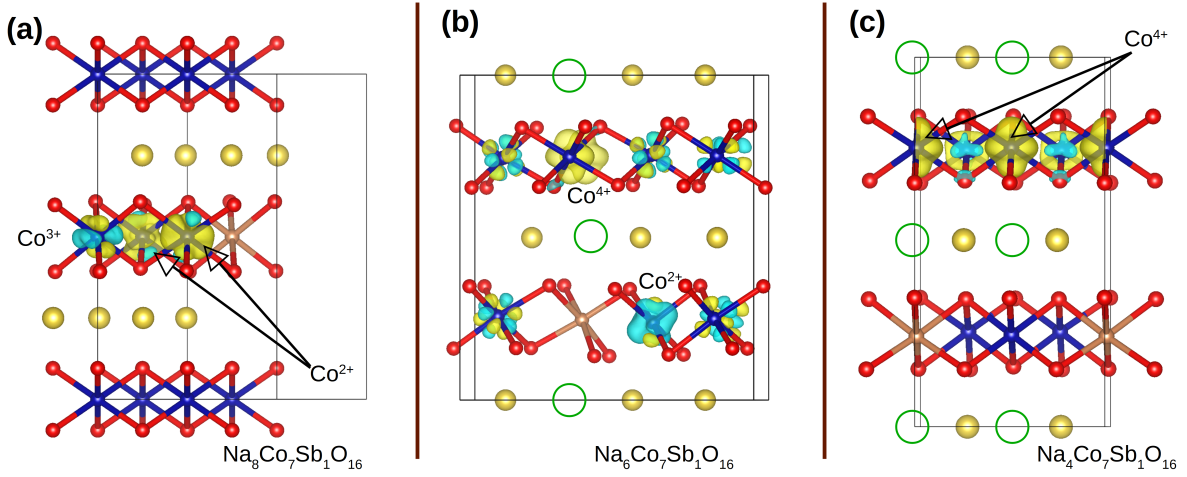

        \begin{minipage}{\textwidth}
            \centering
            \includegraphics[width=0.9\textwidth]{{{Fig2}}}
            \caption{\label{fig:2}Spin densities ($\rho$) at Co sites in Sb doped \ch{Na_{x}CoO2}  structures. Na, O, Co and Sb ions are represented by yellow, red, blue and brown spheres respectively. The green circles denote the sites from which Na ions were vacated. Yellow and cyan iso-surfaces correspond to the up spin and down spins respectively. The iso-surfaces were adown at $\rho = 0.025 e\si{./\angstrom^3}$.}
        \end{minipage}
        \end{figure*}

        The spin density of the ground states of Sb doped \ch{Na_{x}CoO2}  compounds is presented in FIG. \ref{fig:2}. In all compounds, we see that the electrons of the of the \ch{Sb^{5+}} ion localize on the neighboring Co ions. In \ch{Na_{1}CoO2:Sb^{5+}}, there are two \ch{Co^{2+}} ions neighboring \ch{Sb^{5+}}. In \ch{Na_{0.75}CoO2}, a hole created by Na vacancy in the bottom layer neutralizes the electron on one of the \ch{Co^{2+}} ion, nonetheless, the second \ch{Co^{2+}} retains its electron while second Na vacancy from the top layer oxidizes a nearby \ch{Co^{3+}} to \ch{Co^{4+}}. In \ch{Na_{0.5}CoO2:Sb^{5+}}, the two Na vacancies from the bottom layers neutralize the electrons on the two \ch{Co^{2+}} ions neighboring to \ch{Sb^{5+}} while the Na vacancies in the top layer are oxidized by the Na vacancies in the top layer. Furthermore, we see that the magnetic moment bearing Co ions couple ferromagnetically with each other when they are located in the same \ch{CoO2} layer and antiferromagnetically when they are located in different \ch{CoO2} layers. This trend is similar to that of undoped \ch{Na_{x}CoO2}  which exhibits in-plane ferromagnetism and out of plane antiferromagnetism (A-type) for a considerable range of $x$ \cite{Ihara2004, Bayrakci2005, Galeski2016}.

        To examine the strength of magnetic coupling, we recalculated the total energy of the \ch{Sb^{5+}} doped compounds under constraints that fixed the spin coupling to the opposite settings of the ground states presented in FIG. \ref{fig:2}. We found that the total energy of \ch{Na_{1}CoO2:Sb^{5+}}  increased by 61.577 meV/f.u. (f.u. is formula unit) when the two \ch{Co^{2+}} ions were ferromagnetically coupled; the total energy of \ch{Na_{0.75}CoO2:Sb^{5+}}  increased by 68.551 meV/f.u. when the \ch{Co^{2+}} and \ch{Co^{4+}} had parallel spin alignment and the total energy of \ch{Na_{0.5}CoO2:Sb^{5+}}  increased by 65.336 meV/f.u. when the \ch{Co^{4+}} ions in the top layer had antiferromagnetic alignment. Our results demonstrate that similar to the undoped \ch{Na_{x}CoO2}, the in-plane and out-of-plane magnetic couplings are of the same magnitude \cite{Johannes2005}.

        To investigate the electrochemical activity of the Sb doped \ch{Na_{x}CoO2}, we calculated the average voltage of the \ch{Na_{x}CoO2:Sb}  cathode as a function of Na content ($x$) using the following equation:
        \[V = -\frac{\left[E^t \left(\ch{Na_{x}CoO2:Sb} \right) - E^t\left(\ch{Na_{x-y}CoO2:Sb}\right)-E^t\left(\ch{Na}\right)\right]} {\left(x-y\right)e}\]
        in which $E^t\left(\ch{Na_{x}CoO2} \right)$ is the total energy of the \ch{Na_{x}CoO2}  compound, $E^t\left(\ch{Na}\right)$ is the total energy of the elemental solid Na per atom and $e$ is the electronic charge.
        We found that the average potential was 2.140 V for $1 < x< 0.75$ and 2.373 V for $0.75 < x <0.5$. These voltages are slightly smaller than that of the pristine compound for which we calculated voltages of 2.396 V for $1 < x< 0.75$ and 2.880 V for $0.75 < x <0.5$. The slight decrease in the potential may have been caused by \ch{Sb^{5+}} stronger electrostatic repulsion towards Na ions which facilitates easier Na extraction.

        \begin{table}
        \centering
        \caption{\label{tab:2}Number of electron bearing Co ions (\ch{Co^{2+}}) and number of hole bearing Co (\ch{Co^{4+)}} and the relative energy in \ch{Na_{x}CoO2:Sb^{3+}}  and \ch{Na_{x}CoO2:Sb^{5+}}  The relative energy is presented with respect to the most stable compound for a given Na content.}
        \begin{tabularx}{\linewidth}{l*4{p{0.9cm}}X}
            \thickhline{1.5pt}
            System & No. \ch{Co^{2+}} & No. \ch{Co^{3+}} & No. \ch{Co^{4+}} & Total Spin & Relative Energy (eV/f.u.)\\
            \thickhline{1pt}
            \ch{Na8Co7Sb^{5+}O16} & 2 & 5 & 0 & 2 & 0\\
            \ch{Na8Co7Sb^{3+}O16} & 0 & 7 & 0 & 0 & 0.587\\
            \ch{Na6Co7Sb^{5+}O16}(1) & 1 & 5 & 1 & 2 & 0\\
            \ch{Na6Co7Sb^{5+}O16}(2) & 0 & 7 & 0 & 0 & 0.319\\
            \ch{Na6Co7Sb^{3+}O16} & 0 & 5 & 2 & 2 & 0.305\\
            \ch{Na4Co7Sb^{5+}O16} & 0 & 5 & 2 & 2 & 0\\
            \ch{Na4Co7Sb^{3+}O16} & 0 & 3 & 4 & 4 & 0.0115\\
            \thickhline{1.5pt}
        \end{tabularx}
        \end{table}

        Now let's examine how dopants affect the thermoelectric performance of the sodium cobaltate. The Seebeck effect in \ch{Na_{x}CoO2}  is driven by the spin entropy flow between \ch{Co^{3+}}/\ch{Co^{4+}} according to the modified Heike's formula for $S$ \cite{Terasaki2011}:
        \[S = -\frac{k_B}{e} \ln\left[\frac{g\left(\ch{Co^{3+}}\right)}{g\left(\ch{Co^{4+}}\right)} \frac{t}{1-t}\right].\]
        Here $k_B$ is the Boltzmann constant, $e$ is the electron charge, $t$ is the concentration of \ch{Co^{3+}} ions and $g$ is electronic degeneracy of the Co ions. The electronic degeneracy ($g$) of an ion equals to the different possible ways in which electrons can be arranged. Consequently, $g$ is the product of spin degeneracy ($g\ch{_{spin}}$) and orbital ($g\ch{_{orbital}}$) degeneracy: $g = g\ch{_{spin}}\cdot g\ch{_{orbital}}$. $g\ch{_{spin}}$ equals to $2\zeta + 1$ where $\zeta$ is the total spin number and $g\ch{_{orbital}}$  is the number of valid permutations for distributing the electrons across the orbitals. At their low spin states ($\zeta = \tfrac{1}{2}$ for \ch{Co^{2+}}, $\zeta = 0$ for \ch{Co^{3+}} and $\zeta = \tfrac{1}{2}$ for \ch{Co^{4+}}), $g(\ch{Co^{2+}})$ is 4, $g(\ch{Co^{3+}})$ is 1 and $g(\ch{Co^{4+}})$ is 6.
        In \ch{Na_{1}CoO2:Sb^{5+}}, substituting the corresponding g and t values in Eq. 2 yields $S = 40.50 \si{.\mu V/K}$. We should remember that in pristine \ch{Na_{1}CoO2}, since all Co ions are in +3 oxidation state, there is no spin entropy flow and therefore $S = 0$. In \ch{Na_{0.75}CoO2:Sb^{5+}}, the Seebeck effect is driven by the spin entropy flow between \ch{Co^{3+}}/\ch{Co^{4+}} and \ch{Co^{3+}}/\ch{Co^{2+}} ion pairs. Since the $g(\ch{Co^{2+}})$ is smaller than $g(\ch{Co^{4+}})$, one expects that Sb doping in \ch{Na_{0.75}CoO2:Sb}  would decrease $S$ with respect to $S$ in pristine compound. In \ch{Na_{0.50}CoO2:Sb}, \ch{Co^{2+}} ions are no longer present in the compound but \ch{Sb^{5+}} dopant instead greatly decrease the concentration of \ch{Co^{4+}}. Consequently, we obtain an $S$ value of $75.44 \si{.\mu V/K}$ which is less than half of the value of $S$ value in pristine \ch{Na_{0.5}CoO2}  which is $154.40 \si{.\mu V/K}$.

    \section{Conclusions}
        We demonstrated that Sb dopant in \ch{Na_{x}CoO2}  adopt the highest possible oxidation state, in this regard, is acts similar to \ch{Eu^{3+}}, \ch{Ni^{4+}} \cite{Assadi2013}, \ch{Cu^{2+}} \cite{Assadi2015c} and \ch{Ru^{4+}} \cite{Assadi2017} dopants. The electrons generated by \ch{Sb^{5+}} high oxidation state, localize on the neighboring Co ions instead of being delocalized over the valence band. This phenomenon results in the creation of \ch{Co^{2+}} ions which do not occur in \ch{Na_{x}CoO2}. Magnetically, \ch{Co^{2+}} ions behave similarly to \ch{Co^{4+}} as they couple ferromagnetically within the same \ch{CoO2} layer and couple antiferromagnetically across \ch{CoO2} layers with \ch{Co^{4+}} ions. The higher electrostatic repulsion between \ch{Sb^{5+}} and the \ch{Na+} ions facilitates easier \ch{Na+} extraction that eventually results in a slight decrease in the electrochemical potential. We therefore anticipate \ch{W^{6+}}, \ch{Bi^{5+}}, \ch{Sn^{4+}} \cite{Assadi2015c} dopants having similar effect in reducing the Na extraction potential.

    \begin{acknowledgments}
        This work was supported by Australian Research Council. The high performance computational facilities were provided by Intersect Australia Limited and Kyushu Universitys supercomputing center.
    \end{acknowledgments}

    \bibliography{Paper5}{}
\end{document}